\documentstyle[aps,prb,epsf,preprint]{revtex}
\begin{document}



\title{Locally critical quantum phase transitions in strongly correlated metals}
\author{Qimiao Si$^\ast$, Silvio Rabello$^\ast$,
Kevin Ingersent$^\dag$, and J.\ Lleweilun Smith$^\ast$
}
\address{$^\ast$Department of Physics \& Astronomy, Rice University, Houston,
TX 77251--1892, U.S.A. \\
$^\dag$Department of Physics, University of Florida, Gainesville,
FL 32611--8440, U.S.A.}

\maketitle

\vskip 0.5 in

{\bf

When a metal undergoes a continuous quantum phase transition,
non-Fermi liquid behaviour arises near the critical point.
It is standard to assume that all low-energy degrees of freedom induced by
quantum criticality are spatially extended,
corresponding to long-wavelength fluctuations of the order parameter.
However, this picture has been contradicted by recent experiments on a
prototype system: heavy fermion metals at a zero-temperature magnetic
transition.
In particular, neutron scattering from CeCu$_{6-x}$Au$_x$ has revealed
anomalous dynamics at atomic length scales, leading to much debate
as to the fate of the local moments in the quantum-critical regime.
Here we report our theoretical finding of a locally critical quantum
phase transition in a model of heavy fermions.
The dynamics at the critical point are in agreement with experiment.
We also argue that local criticality is a phenomenon of general relevance to
strongly correlated metals, including doped Mott insulators.

}


Quantum (zero-temperature) phase transitions
are ubiquitous in strongly correlated metals; for
a recent review,
see ref.~\onlinecite{Sachdev-book}.
The extensive current interest in metals close to a second-order
quantum phase transition has stemmed largely from studies of
high-temperature superconductors. In these systems, however,
it has been hard to actually locate the putative quantum
critical points (QCPs).
The situation appears to be simpler
in some related families of strongly correlated metals. In particular,
there are many heavy fermion metals which can be tuned between
an antiferromagnetic (AF) metal and a paramagnetic metal.
In recent years QCPs have been explicitly identified in a number of
stoichiometric or nearly stoichiometric systems.
Non-Fermi liquid behaviours\cite{Maple} are observed, usually through
transport and thermodynamic measurements, in the quantum critical regime.
Examples\cite{Lonzarich,vonLohneysen,Steglich,Stewart,%
deVisser,RMP,Schroder1,Stockert,Schroder2}
include CePd$_2$Si$_2$, CeCu$_{6-x}$Au$_x$, CeNi$_2$Ge$_2$, and YbRh$_2$Si$_2$.

The most fundamental question regarding any condensed matter system
concerns the nature of their low-energy excitations.
The traditional theory for metals close to a QCP assumes that the only
important degrees of freedom
are long-wavelength fluctuations of the order parameter.
For a magnetic QCP,
such long-wavelength fluctuations are usually called paramagnons.
This theory has its origins back in the 1960s in the description
of liquid helium-3; for a historical perspective,
see ref.~\onlinecite{Lonzarich-review}. It was formulated 
in the modern language of critical phenomena
in refs.~\onlinecite{Hertz} and~\onlinecite{Millis}.
The assumption that the long-wavelength paramagnons are the only
critical degrees of freedom, adopted from the theory of classical
(finite-temperature) phase transitions, 
turns out to have 
three important consequences.
Firstly, the dynamical spin
susceptibility at generic wavevectors (wavevectors far
from the AF ordering wavevector) has the usual Fermi-liquid form,
i.e., it is linear in frequency.
Secondly, even close to the ordering wavevector, the dynamical
spin susceptibility is also linear in frequency.
Thirdly, there is a violation of the so-called $\omega/T$ scaling
(defined in the next section).

This picture has been subjected to experimental test in recent years.
Heavy fermion metals have undergone by far the most systematic study.
A major puzzle has been raised by neutron-scattering
experiments\cite{Schroder1,Stockert,Schroder2} on CeCu$_{6-x}$Au$_x$,
which goes from a paramagnetic metal to an antiferromagnetic metal as
$x$ increases through a critical value, $x_c \approx 0.1$
(ref.~\onlinecite{vonLohneysen}).
Close to the AF ordering wavevector, a fractional exponent
($\alpha < 1$) appears in the frequency and temperature dependences
of the dynamical spin susceptibility.  The same exponent $\alpha$
describes the frequency and temperature dependences
at generic wavevectors, over essentially the entire Brillouin zone.
Finally, the dynamical spin susceptibility exhibits $\omega/T$ scaling.
These experiments have lead to much debate as to the nature of the quantum
critical points in heavy fermion metals, especially concerning the fate
of the local magnetic moments in the quantum-critical regime\cite{Schroder1,%
Stockert,Schroder2,Coleman,SiSmithIngersent,Rosch,Coleman2}.

Here we show that a dynamical spin susceptibility of the form observed
experimentally may arise at a {\it locally critical quantum phase transition},
where local critical degrees of freedom co-exist with spatially
extended ones. While we reach this conclusion by solving
a microscopic model appropriate for heavy fermion metals,
the possibility of such a novel QCP can be seen on general grounds.
The universality class of a classical phase transition is
entirely determined by statics\cite{Sachdev-book}.
The critical degrees of freedom must be spatially extended, as
only long-wavelength fluctuations of the order parameter are
energetically favourable. At a quantum phase transition, by contrast,
both the statics and dynamics (i.e., quantum fluctuations)
are important\cite{Sachdev-book}.
The effect of quantum fluctuations on local degrees of freedom
is to introduce a coupling to certain dissipative baths.
When this coupling is sufficiently strong, the local degrees of freedom
can become critical.
(A similar situation---local degrees of freedom driven critical by a
dissipative bath---arises in macroscopic quantum tunneling
problems\cite{Leggett}.)

\vskip 0.3 in

\noindent{\bf Locally critical quantum phase transition}

The microscopic model we consider is the Kondo lattice model\cite{Hewson},
illustrated in Fig.~\ref{fig1}a.
At each lattice site, a local moment\cite{Anderson61} interacts via an
exchange coupling $J_K$ with the spin of any conduction electron sitting
at the site.
There are two important energy scales in the problem\cite{Doniach,Varma}:
The Kondo temperature $T_K$ sets the scale below which an isolated local moment
would be screened by the spins of the conduction electrons, while the RKKY
interaction characterizes the induced coupling between two local moments.
We will assume an average occupancy of less than one conduction electron per
lattice site;
as a result, all phases considered here are metallic.
To systematically analyze the behaviour of the local degrees of freedom
in this model, we apply the extended dynamical mean-field theory (EDMFT)
developed in refs.~\onlinecite{Smith1,Smith2,Chitra}.
As in the standard dynamical mean-field theory\cite{Georges},
the correlation functions of the lattice problem in the EDMFT
are calculated through a self-consistent impurity Kondo problem.
The latter, illustrated
in Fig.~\ref{fig1}b, describes a local moment simultaneously coupled to
two dissipative baths. A fermionic bath accounts for all temporal fluctuations
arising from hopping of electrons between the local site and the rest of the
lattice, while a bosonic bath represents the
fluctuating magnetic field generated by the local moments at all other
sites. The couplings to the two baths are $J_K$ and $g$, respectively.
Further details of the method are described later in the paper.

We find two types of QCP, illustrated by the phase diagrams in Figs.~\ref{fig2}
and~\ref{fig3}.
In each case, the tuning parameter $\delta$ is the ratio of the RKKY
interaction to the Kondo temperature.
Increasing $\delta$ has two effects.
First, the magnetic correlations become more pronounced.
The dynamical spin susceptibility $\chi ({\bf Q}, \omega)$---where
${\bf Q}$ is the peak wavevector---diverges at some threshold value
$\delta=\delta_c$, which
defines a QCP separating a paramagnetic metal ($\delta<\delta_c$)
from an antiferromagnetic metal ($\delta > \delta_c$).
Second, the local Kondo physics governed by the effective impurity
problem is also changed\cite{Smith3,Sengupta,SachdevYe}.
Through self-consistency, an increase in $\delta$ causes an
increase in the coupling $g$ between the local moment and the
fluctuating magnetic field.
At some value
$\delta=\delta_{\rm loc}^c$, the corresponding ratio $g/T_K$
reaches a threshold value where the local Kondo problem becomes critical.
The existence of such a critical point reflects
two competing processes in the local problem (Fig.~\ref{fig1}b):
quenching of the local moment through its
Kondo coupling to the spins of the fermionic bath, and the
tendency of the coupling $g$ to polarize the local moment along the direction
of the fluctuating magnetic field.
At $\delta=\delta_{\rm loc}^c$, the fluctuating magnetic field just
succeeds in preventing the spins of the fermionic bath from completely
quenching the local moment, yielding a singular local susceptibility.
This critical point is also marked by the vanishing of a
local energy scale $E_{\rm loc}^*$, reflecting a critical slowing down.
$E_{\rm loc}^*$ serves as an effective Fermi energy scale, in the sense that
the local susceptibility has the usual Fermi liquid form (i.e., an imaginary
part that is linear in $\omega$) only at energies below $E_{\rm loc}^*$.

At the first type of QCP, shown in Fig.~\ref{fig2},
the lattice system orders before the effective
local problem has a chance to become critical, i.e.,
$\delta_c < \delta_{\rm loc}^c$, so $E_{\rm loc}^*$ remains finite at the
transition.
The local moments are completely quenched at zero
temperature over the entire paramagnetic region and also at the QCP.
Their only vestiges are the Kondo resonances,
which hybridize with the conduction electrons to form
Landau quasiparticles \cite{Doniach,Varma,Hewson,Continentino}.
The RKKY interaction manifests itself as an interaction between
these quasiparticles, leading to a spin-density-wave (SDW) instability which
drives the transition \cite{Doniach}.
As a result, the dynamical spin susceptibility has the same form as in the
traditional theory \cite{Hertz,Millis}. We find this type of
QCP when the spin fluctuations are three-dimensional.

The second type of QCP is illustrated in Fig.~\ref{fig3}.
Here, $\delta_c = \delta_{\rm loc}^c$, so the lattice system
reaches its ordering transition precisely at the point where the local problem
also becomes critical. Thus, $E_{\rm loc}^*$ vanishes at the QCP, and two
kinds of critical degrees of freedom
co-exist: long-wavelength fluctuations of the order parameter, and local
fluctuations originating from the local moments. The transition is
{\em locally critical}.
We find this type of QCP when spin fluctuations are two-dimensional.
(In order for the ordering temperature $T_N$ shown in Fig.~\ref{fig3} to be
nonzero for $\delta > \delta_c$, it is necessary to have an infinitesimal
RKKY coupling in the third dimension.)

At the locally critical point, the zero-temperature
dynamical spin susceptibility
has an anomalous frequency dependence at wavevectors $\bf q$ not only close
to the ordering wavevector ${\bf Q}$ but everywhere else in the Brillouin
zone as well:
\begin{equation}
 \chi({\bf q}, \omega) = {1 \over {f({\bf q})
+ A \,\omega^{\alpha}}}.
\label{chi-qw-T=0}
\end{equation}
Here $\alpha$ is an anomalous exponent [an expression for which is
given in equation~(\ref{alpha}) below].
For any wavevector ${\bf q}$ far away from ${\bf Q}$, $f({\bf q})$ is
non-zero and varies smoothly with ${\bf q}$; for ${\bf q}$ close to
${\bf Q}$, $f({\bf q})$ is proportional to $({\bf q} - {\bf Q})^2$.

At finite temperatures,
$\omega^{\alpha}$ in equation~(\ref{chi-qw-T=0}) is replaced by
$T^{\alpha} {\cal M}(\omega/T)$, where ${\cal M}(\omega/T)$ is a scaling
function whose form depends only on $\alpha$.
This has two experimentally testable consequences.
First, the dynamical spin susceptibility at ${\bf q}={\bf Q}$ satisfies
an $\omega/T$ scaling:
\begin{equation}
 \chi({\bf Q}, \omega, T) = {1 \over A T^{\alpha}
 {\cal M}(\omega/T)} .
\label{w/T}
\end{equation}
Second, the
static uniform spin susceptibility has a modified Curie-Weiss form
\begin{equation}
 \chi(T) = {1 \over {\Theta + B \, T^{\alpha}}} ,
\label{chi(T)}
\end{equation}
where $\Theta$ is a positive constant.

Our results can be directly compared to neutron-scattering
experiments on Au-doped CeCu$_{6-x}$Au$_x$ at the critical
concentration $x_c \approx 0.1$:
(i)~The experimental data have been fitted \cite{Schroder1,Schroder2} to the
form of equation~(\ref{chi-qw-T=0}), with an exponent $\alpha \approx 0.75$.
(ii)~Within experimental resolution,
$f({\bf q})$ goes to zero along lines in the three-dimensional
Brillouin zone, implying that the magnetic fluctuations are
two-dimensional in real space \cite{Stockert,Schroder2,Rosch-neutron}.
(iii)~An $\omega/T$ scaling
has been extensively reported \cite{Schroder1,Schroder2}
(see also ref.~\onlinecite{Aronson}).

In addition, the modified Curie-Weiss form [equation~(\ref{chi(T)})] is
known\cite{Schroder1,Schroder2} to fit the magnetization data in
CeCu$_{6-x}$Au$_x$.
This form also appears to describe the magnetization of
some other heavy fermion metals near quantum criticality\cite{RMP}.
One example\cite{Steglich-priv} is YbRh$_2$Si$_2$,
which is already very close to being critical at ambient pressure
and without any doping.  The thermodynamic and transport
properties\cite{Steglich} of this compound are very similar
to those of CeCu$_{6-x}$Au$_x$ at
$x=x_c$.

From the local susceptibility
we can also determine the NMR
spin-lattice relaxation rate $1/T_1$.
Barring subtle form-factor cancellations, the
relaxation rate is expected to be temperature-independent
in the relevant temperature range:
\begin{equation}
{1 \over T_1} \sim {\rm constant}.
\label{T1}
\end{equation}
This prediction could be tested, for instance, through NMR measurements on the
Cu sites in CeCu$_{6-x}$Au$_x$ or the Si sites in YbRh$_2$Si$_2$.

\vskip 0.3 in

\noindent{\bf Extended dynamical mean field theory}

We now outline the EDMFT analysis leading to the conclusions reported above.
The Kondo lattice model illustrated in Fig.~\ref{fig1}a is
specified by the Hamiltonian
\begin{equation}
{\cal H}
= \sum_{ ij,\sigma} t_{ij} ~c_{i\sigma}^{\dagger} ~c_{j\sigma}
+ \sum_i J_K ~{\bf S}_{i} \cdot {\bf s}_{c,i}
+ \sum_{ ij} I_{ij} ~{\bf S}_{i} \cdot {\bf S}_{j} ,
\label{kondo-lattice}
\end{equation}
where ${\bf s}_{c,i}$ is the conduction electron spin at site $i$.
Within the EDMFT\cite{Smith1,Smith2,Chitra},
equation~(\ref{kondo-lattice}) is mapped onto an effective single-site
problem,
illustrated in Fig.~\ref{fig1}b and represented by the Hamiltonian
\begin{eqnarray}
{\cal H}_{\rm loc}
&=& J_K ~{\bf S} \cdot {\bf s}_c
+ \sum_{p,\sigma} E_{p}~c_{p\sigma}^{\dagger}~ c_{p\sigma}
+ \; g \sum_{p} {\bf S} \cdot
\left( \vec{\phi}_{p} + \vec{\phi}_{-p}^{\;\dagger} \right)
+ \sum_{p} w_{p}\,\vec{\phi}_{p}^{\;\dagger}\cdot \vec{\phi}_{p} ,
\label{H-imp}
\end{eqnarray}
where $c_{p\sigma}$ and $\vec{\phi}_{p}$ describe the fermionic
and bosonic dissipative baths, respectively.
The self-consistent procedure goes as follows.
We put in trial forms for $E_{p}$, $w_{p}$, and $g$, and solve the impurity
Kondo problem to determine the electron self-energy $\Sigma(\omega)$,
the ``spin self-energy'' $M(\omega)$, the local Green's function
$G_{\rm loc}(\omega)$, and the local susceptibility $\chi_{\rm loc}(\omega)$.
(For definitions of these functions and further details of the calculational
procedure, see the Methods section. We note here that the key approximation
of the EDMFT is the assumption that the electron and spin self-energies
are momentum-independent.)
Self-consistency is imposed by demanding that $G_{\rm loc}$ and
$\chi_{\rm loc}$ are equal to the wavevector averages, respectively, of the
lattice Green's function $G({\bf k},\omega)$ and the lattice susceptibility
\begin{equation}
\chi ({\bf q}, \omega) = \frac {1}  { M(\omega) + I_{\bf q} } ,
\label{chi-def}
\end{equation}
where $I_{\bf q}$ is the Fourier transform of $I_{ij}$.
For the purpose of determining the universal low-energy behaviour,
we take the density of states of the fermionic
bath near the Fermi energy to be a non-zero constant,
and the spectral function of the fluctuating magnetic field to have a
power-law dependence on frequency, with an exponent $\gamma$,
for $\omega$ below some cut-off scale $\Lambda$.
The effective impurity Kondo problem is solved using a $1-\gamma$
expansion\cite{Smith3,Sengupta,Vojta}.
Two types of self-consistent solution are obtained for different forms of the
``RKKY density of states'',
\begin{equation}
\rho_{I} (\epsilon) = \sum_{\bf q} \delta ( \epsilon  - I_{\bf q} ).
\label{rkky-dos}
\end{equation}

The first type of solution, illustrated by Fig.~\ref{fig2}, arises
when the RKKY density of states $\rho_{I}(\epsilon)$
increases from the lower band edge
(at $\epsilon = I_{\bf Q}$)
in a square-root fashion
(i.e., as $\sqrt{\epsilon - I_{\bf Q}}\,$).
At the QCP, the spin self-energy takes the form
\begin{equation}
M (\omega ) = -I_{\bf Q} - i \, a \, \omega ,
\label{M-sdw}
\end{equation}
where $a$ is a positive, real constant.
Through equation~(\ref{chi-def}), this yields a
$\chi({\bf q},\omega)$ of the form given in the traditional
theory \cite{Hertz,Millis}.
The linear frequency dependence of $M(\omega)$
reflects the physics that the
spin fluctuations actually describe quasiparticle-quasihole
pairs.

The second type of solution, illustrated by Fig.~\ref{fig3}, arises when
$\rho_{I}(\epsilon)$
increases from zero at the lower band edge with a jump.
The self-consistent local dynamical spin susceptibility at the QCP
is singular, and is given by
\begin{equation}
\chi_{\rm loc} (\omega) = { 1 \over {2 \Lambda}}
                ~\ln {\Lambda \over {- i \omega}} ,
\label{chi-loc-crit}
\end{equation}
where
\begin{equation}
\Lambda = { 2 \over {\pi \rho_0 (\mu)}}
\exp \left[- {1 \over \rho_0 (\mu)J_K }\right] ,
\label{Lambda}
\end{equation}
$\rho_0 (\mu)$ being the bare conduction electron density of states
at the chemical potential $\mu$.
The corresponding spin self-energy has the form
\begin{equation}
        M (\omega ) ~\approx -I_{\bf Q} + A ~\omega^{\alpha}.
\label{M-loc-crit}
\end{equation}
Here, $A = (-i)^{\alpha} \Lambda_0 \Lambda^{- \alpha}$,
where $\Lambda_0$ is defined by the condition that,
for $\epsilon \in (I_{\bf Q}, I_{\bf Q}+\Lambda_0)$,
$\rho_I (\epsilon)$ is approximately equal to its value
at the lower edge, $\rho_I (I_{\bf Q})$.
The exponent $\alpha$
depends on both $\Lambda$
and $\rho_{I}(I_{\bf Q})$:
\begin{equation}
\alpha = { 1 \over {2 \Lambda \rho_{I}(I_{\bf Q}) }} .
\label{alpha}
\end{equation}
The spin self-energy acquires this anomalous frequency dependence because
spin fluctuations can not only decay into particle-hole pairs, but can
also couple to critical local modes.
The dynamical spin susceptibility at zero temperature is
then given by equation~(\ref{chi-qw-T=0}),
with
$f({\bf q}) = I_{\bf q} - I _ {\bf Q}$.
At finite temperatures,
the self-consistent solution gives
a spin self-energy
\begin{equation}
        M (\omega, T ) ~\approx -I_{\bf Q} +
A ~T^{\alpha} {\cal M}(\omega/T).
\label{M-loc-crit-finite-T}
\end{equation}
The scaling function
is ${\cal M}(\omega/T) = (i 2 \pi)^{\alpha} \exp[\alpha~ \psi
(1/2 - i \omega / 2 \pi T)]$,
where $\psi$ is the digamma function.
This directly leads to equations~(\ref{w/T}) and~(\ref{chi(T)}),
with $\Theta = I_{{\bf q}={\bf 0}} - I_{\bf Q}$ and
$B \approx \Lambda_0\Lambda^{-\alpha}
(2 \pi)^{\alpha}\exp[\alpha \psi(1/2) ]$.

A square-root onset and a jump at the lower edge of the RKKY density of states
are characteristic of magnetic fluctuations in three and two dimensions,
respectively.
These behaviours are exemplified by the RKKY density of states associated
with nearest-neighbor coupling in a (three-dimensional) cubic lattice
and its counterpart in a (two-dimensional) square lattice.
While the details depend on the lattice type and the range of interactions,
the square-root and jump onsets in the two cases are robust
provided that $I_{\bf q}$ approaches $I_{\bf Q}$ in a $({\bf q} - {\bf Q})^2$
fashion.

\vskip 0.3 in

\noindent{\bf Beyond the microscopic theory}

So far we have presented the results of a
specific
microscopic approach:
the extended dynamical mean field theory
for the Kondo lattice model.
We now turn to more general considerations concerning the
existence and properties of locally critical quantum phase transitions.
There are four main points that we wish to discuss.

Firstly, even in the traditional theory there are hints, albeit very subtle,
that two-dimensionality facilitates the realization of a
locally critical point. In this theory \cite{Hertz,Millis},
the zero-temperature spin fluctuations are given by the harmonic,
or Gaussian, fluctuations of the order parameter.
As a result, the zero-temperature, zero-frequency spin
susceptibility is proportional to $({\bf q} - {\bf Q})^{-2}$.
In two dimensions, the corresponding
local---i.e., wavevector-averaged---susceptibility is in fact logarithmically
singular.
(For two-dimensional fermions coupled to two-dimensional commensurate
spin fluctuations, there are other kinds of singularities
in the SDW description; see ref.~\onlinecite{Chubukov}.)
This singularity can cause important non-linear effects in the dynamics
of any non-trivial local degrees of freedom---such as local moments in
heavy fermion metals---and can ultimately drive them critical.
Our EDMFT analysis amounts to a microscopic prescription for
these non-linear effects. We note that the situation is very different
for quantum critical points in
insulators\cite{Chakravarty,Sachdev-book}.
There the effective field theory is usually below its upper critical
dimension, implying that the spin susceptibility is
$({\bf q} - {\bf Q})^{-2+\eta}$, where the anomalous exponent
$\eta$ takes a positive value. The corresponding local
susceptibility is no longer singular in two dimensions.

Secondly, the contrast between a locally critical transition
and an SDW transition can also be seen at the level of a
Ginzburg-Landau-like
description. For an SDW transition,
there is only one type of critical degrees of freedom,
namely the long-wavelength fluctuations of the order parameter.
The Ginzburg-Landau (GL)
action is then written purely in terms of
the frequency-dependent magnetization at wavevectors close to
the ordering wavevector:
\begin{equation}
{\cal S}_{\rm SDW} =
{\cal S}_{\rm lw}\,[\, {\bf m} ({\bf q} \sim {\bf Q}, \omega ) \,] .
\label{GL-sdw}
\end{equation}
This generalizes the standard $\phi^4$ theory for a classical phase
transition in that the fluctuations take place not only in the $D$
spatial dimensions, but also in time.
The temporal fluctuations can be thought of as adding $z$ dimensions,
where $z$ is the dynamical exponent:
${\cal S}_{\rm lw}$ is a $\phi^4$ theory with
an effective dimensionality $D_{\rm eff}=D+z$;
see refs.~\onlinecite{Hertz,Millis,Sachdev-book}.

For a locally critical point, on the other hand, the GL action
is no longer just a $\phi^4$ theory.
There are extra critical modes which characterize the continuous
disappearance of the Kondo resonance as the critical point
is approached from the paramagnetic side.
These modes are independent of the fluctuations of the order parameter.
They are spatially local since the destruction of the Kondo resonance
is a local phenomenon.
The GL action will now
contain, in addition to the $\phi^4$  component, ${\cal S}_{\rm lw}$,
parts describing these new local critical degrees of freedom and their
non-linear coupling to the long-wavelength modes.
The explicit construction of these additional parts of the GL action
is left for future work.

Thirdly, consideration of a GL description
makes it likely that the
locally critical point is stable beyond our EDMFT approximation,
provided that $\alpha < 1$
(as is the case
in CeCu$_{6-x}$Au$_x$ and ${\rm Yb Rh_2Si_2}$).
The crucial question is
what happens when we allow an explicit ${\bf q}$ dependence in
the spin self-energy.
The dynamical exponent entering ${\cal S}_{\rm lw}$ will be $z = 2/\alpha$.
When $\alpha < 1$, the effective dimension $D_{\rm eff}=D+z=2+2/\alpha$
will be
above the upper critical dimension of
4, so that
all the non-linear
couplings within ${\cal S}_{\rm lw}$
will be
irrelevant in the renormalization-group
sense \cite{Sachdev-book}. As a result,
the ${\bf q}$-dependence of the spin  self-energy from ${\cal S}_{\rm lw}$
will be at most $({\bf q}- {\bf Q})^2$.
We also expect that
the spin self-energy due to the coupling of the order-parameter field
with the local modes will have a smooth ${\bf q}$-dependence.
As a result, the zero-temperature spin
susceptibility
should be
proportional to $({\bf q} - {\bf Q})^{-2}$
(i.e., the anomalous exponent $\eta = 0$).
This implies that the singular local susceptibility
and the
concomitant
destruction of the Kondo resonance at the critical point are
robust in two dimensions.
It should be noted that a rigorous proof of this stability awaits
the explicit construction of the GL action for the locally
critical point.

Fourthly,
GL considerations also
provide a way
to understand the $\omega/T$ scaling [equation~(\ref{w/T})]
at the locally critical point.
At an SDW transition, the field theory, equation~(\ref{GL-sdw}),
is a $\phi^4$ theory with an effective dimensionality
$D_{\rm eff} = d+z$.
Whereas at QCPs in insulators\cite{Chakravarty,Sachdev-book} the dynamical
exponent is usually $1$, in a metallic environment $z=2$, raising
$D_{\rm eff}$ above the upper critical dimension of~4 for
the $\phi^4$ theory. All the non-linear couplings are then
irrelevant, and $\omega/T$ scaling is absent\cite{Sachdev-book,Millis}.
(A violation of $\omega/T$ scaling has been reported\cite{Raymond}
near a QCP in ${\rm Ce_{1-x}La_xRu_2Si_2}$.)

At a locally critical point, the non-linear couplings among the
long-wavelength modes remain irrelevant.  However, the GL action
also contains non-linear couplings that involve the additional
local critical modes. These couplings are relevant, making the
field theory an interacting one, thereby \cite{Sachdev} allowing
the $\omega/T$ scaling given in equation~(\ref{w/T}).
In terms of a suitably defined
spin relaxation rate, the contribution from the
$\phi^4$ component depends on temperature with a power greater
than~1. On the other hand, the contribution from the coupling
to the local modes is linear in temperature due to the ``relevant''
non-linear couplings.
The total relaxation rate is then linear in temperature,
as required\cite{MFL} for
$\omega/T$ scaling. Again, the EDMFT analysis of the previous section
amounts to a microscopic prescription for determining the relaxation
rate and, more generally, the entire scaling function.

\vskip 0.3 in

\noindent{\bf Broader implications}

In addition to explaining the salient features of
experiments on
heavy fermion metals close to a quantum critical point,
our results may have broader implications for other
strongly correlated metals.
The emergence of critical local modes in the case we have
studied
depends crucially on the formation of local moments.
A local moment is nothing other than a strongly interacting $f$-electron
orbital which cannot be doubly occupied due to a strong on-site
Coulomb interaction\cite{Anderson61}. It is formed
at energies intermediate between the bare Coulomb interaction scale and
the asymptotic low-energy limit.
The existence of non-trivial local physics at such intermediate
energy scales is in fact
ubiquitous in strongly correlated metals,
including metals close to a Mott insulator.
This suggests the possibility of a locally critical point
occurring in a doped Mott insulator.

\vskip 0.3 in

\noindent{\bf Methods}

We adopt the EDMFT as a conserving resummation of diagrams
for finite dimensional systems [Fig.~7(b) of ref.~\onlinecite{Smith1}].
The spatial dimensionality enters, among other ways, via
the form of the
RKKY density of states defined in equation~(\ref{rkky-dos}).
For any finite-dimensional system, the support to
$\rho_{I} (\epsilon)$ is bounded and
a stable paramagnetic solution exists.

The EDMFT provides a self-consistent procedure. In the
Bose-Fermi Kondo Hamiltonian, equation~(\ref{H-imp}),
we insert trial forms for the spectra of both the vector-bosonic bath
and fermionic bath, which can be parameterized in terms of the Weiss fields
$\chi_0^{-1}$ and $G_0^{-1}$, respectively:
\begin{equation}
g^{2} \sum_{p}  { 2 w_{p} \over {\omega^{2} - w_{p}^{2}}}
= - \chi_{0}^{-1}(\omega),
\quad
\sum_{p} { 1 \over {\omega - E_p}} = G_0(\omega).
\label{weiss-fields-def}
\end{equation}
The trial forms for the QCP are taken to be
\begin{eqnarray}
{\rm Im}\, G_0 (\omega + i 0^+) &=& - \pi N_0,
\nonumber \\[-2ex]
\label{dos-fermion-boson} \\[-2ex]
{\rm Im}\, \chi_0^{-1} (\omega + i 0^{+})
&=& C |\omega|^{\gamma} {\rm sgn}\,\omega
\quad {\rm for} ~ |\omega| < \Lambda,
\nonumber
\end{eqnarray}
where $N_0$, $\gamma$, $C$, and $\Lambda$ are parameters to be
determined.

Next, we solve the Bose-Fermi Kondo problem
using a $(1-\gamma)$ expansion.
This step determines
the local spin susceptibility and the local
conduction electron Green's function,
\begin{eqnarray}
\chi_{\rm loc} (\tau) &=&  - \langle T_{\tau} S_x (\tau) S_x (0)
\rangle_{{\cal H}_{\rm loc}},
\nonumber \\[-2ex]
\label{chi-G-loc} \\[-2ex]
G_{\rm loc} (\tau) &=&
-\langle T_{\tau} c_{\sigma} (\tau) c_{\sigma}^{\dagger} (0)
\rangle_{{\cal H}_{\rm loc}} , \nonumber
\end{eqnarray}
where $\tau$ is the imaginary time. This step also
specifies\cite{Smith1} the spin self-energy $M(\omega)$,
and the conduction electron
self-energy $\Sigma (\omega)$:
\begin{eqnarray}
M( \omega ) &=& \chi_{0}^{-1}( \omega ) + {1 \over \chi_{\text{loc}}
(\omega) },
\nonumber \\[-1ex]
\label{spin-electron-se} \\[-1ex]
\Sigma ( \omega) &=& G
_{0}^{-1}( \omega) - {1 \over G_{\text{loc}}( \omega)} .
\nonumber
\end{eqnarray}
(The spin self-energy is defined in terms of an effective spin
cumulant that is $I$-irreducible\cite{Smith1}.)

The $(1-\gamma)$ expansion we use follows the work of
Smith and Si\cite{Smith3},
Sengupta\cite{Sengupta}, and Vojta
{\it et al.}\cite{Vojta}.
To linear order in $1-\gamma$, there is
a renormalization-group
fixed point located at
$J_K = C = (1-\gamma)/2$.
In analogy to the standard application of the
$\epsilon$ expansion\cite{Wilson-Kogut}, we assume
that the local spin susceptibility $\chi_{\rm loc} (\tau )$ takes
a power-law form at the critical point. By setting $J_K$ and $C$
to their fixed point values, we can determine the exponent by matching the
leading $\ln (\tau)$ term in the expansion of the assumed power-law
form for $\chi_{\rm loc} (\tau )$ with the leading $\ln(\tau)$ term
of the perturbative expression for
$\chi_{\rm loc} (\tau )$, calculated from the
Bose-Fermi Kondo Hamiltonian, equation~(\ref{H-imp}),
to
linear order in $1-\gamma$. This gives,
for $0 < \gamma < 1$,
\begin{equation}
\chi_{\rm loc} (\omega ) = { 1 \over 2 \Lambda^{1-\gamma}}
\Gamma(\gamma) \sin{{\pi (1 - \gamma)}
\over 2} (-i\omega)^{-\gamma} ,
\label{chi-loc-epsilon-omega-gamma}
\end{equation}
where $\Gamma$ is the gamma function, and for
$\gamma = 0$, equation~(\ref{chi-loc-crit}).

In addition, to linear order in $1-\gamma$, the conduction electron
self-energy vanishes:
\begin{equation}
\Sigma = O\left((1-\gamma)^2\right).
\label{Sigma-c-qcp}
\end{equation}

Finally, we demand self-consistency, which amounts to the
requirement that a local correlation function is equal to the
wavevector average of the corresponding lattice
correlation function. In terms of the RKKY density
of states, $\rho_{I} (\epsilon)$
defined in equation~(\ref{rkky-dos}), and the usual
conduction electron density of states, $\rho_0(\epsilon)$,
the self-consistency condition can be written as
\begin{eqnarray}
\chi_{\rm loc} (\omega) &=&
\int d \epsilon
\frac {\rho_I (\epsilon )}
{ M(\omega) + \epsilon } ,
\nonumber \\[-1ex]
\label{self-consistent} \\[-1ex]
G_{\rm loc} (\omega) &=&
\int d \epsilon
\frac {\rho_0 (\epsilon )}
{\omega - \epsilon - \Sigma (\omega)} .
\nonumber
\end{eqnarray}

When $\rho_I (\epsilon)$ is proportional to $\sqrt{\epsilon-I_{\bf Q}}$
near its lower band edge ($\epsilon \rightarrow I_{\bf Q}^+$),
$\chi_{\rm loc}$ is finite at the QCP. The local moments are not critical.

The locally critical point arises when $\rho_I (\epsilon)$
has a jump at the lower band edge.
$\chi_{\rm loc}$ is now singular at the QCP.
The self-consistent solution for the parameters
introduced in equations~(\ref{dos-fermion-boson}) is
\begin{equation}
N_0 = \rho_0 (\mu) , \quad
\gamma = 0^+ , \quad
C = \pi \Lambda,
\label{solution-2D-parameters}
\end{equation}
with the expression for $\Lambda$
being
given in the main text, as are those for
$\chi_{\rm loc}$ and $M$.
We also note that the relationship between $\gamma$ and $\alpha$ is
indirect: it is specified by
the self-consistency equation~(\ref{self-consistent}) along with
equations~(\ref{dos-fermion-boson})--(\ref{spin-electron-se}).

The local susceptibility, given in equation~(\ref{chi-loc-crit}),
is universal. The exponent $\alpha$ for the spin self-energy
depends on the product $\Lambda \rho_I (I_{\bf Q} )$
[see equation~(\ref{alpha})].
It can be seen from equation~(\ref{Lambda}) that $\Lambda$ is of the order
of the Kondo energy.
Since the QCP is reached through a competition
between the RKKY and Kondo interactions, we expect that this product
is close to unity, resulting in an $\alpha$ that is not too far
from ${1 \over 2}$. The precise value of $\alpha$, however,
is interaction-dependent: it depends on which point of the phase
boundary in the RKKY interaction-Kondo energy
parameter space is crossed as the system is tuned through the
quantum phase transition.

At finite temperatures, the self-consistent procedure outlined
above can also be carried through, leading to
the spin self-energy quoted in the main text.

\acknowledgments

We thank G.\ Aeppli, A.\ Chubukov, P.\ Coleman, A.\ J.\ Millis,
A.\ Schr\"oder, A.\ M.\ Sengupta, C.\ M.\ Varma, and P.\ W\"olfle
for useful discussions. This work has been supported by NSF,
TCSUH,
and the A.\ P.\ Sloan Foundation.

Correspondence and requests for materials should be addressed to
Q.\ S. (e-mail: qmsi@rice.edu).

\newpage

\begin{figure}
\caption{
A theoretical model of heavy fermions.
{\bf a},~The Kondo lattice model.
At each lattice site $i$, a spin-${1 \over 2}$ local moment
${\bf S}_{i}$ interacts with a local conduction electron orbital
$c_{i\sigma}$
through an antiferromagnetic Kondo coupling of strength $J_K$.
The hopping amplitude between the conduction electron
orbitals at sites $i$ and $j$ is $t_{ij}$.
The Fourier transform of $t_{ij}$ gives the band dispersion $\epsilon_{\bf k}$,
corresponding to which is a density of states $\rho_0(\epsilon)$.
The Kondo temperature $T_K$ is the characteristic scale for
the screening of an isolated local moment by the spins of the conduction
electrons. For sufficiently small $J_K$,
$T_K \approx [\rho_0 (\mu)]^{-1} \exp[-1/\rho_0(\mu)J_K]$,
where $\mu$ is the chemical potential.
The RKKY interaction between two local moments at sites $i$ and $j$
is $I_{ij}$, the Fourier transform of which is $I_{\bf q}$.
{\bf b},~The effective impurity Kondo model to which the Kondo lattice model
is mapped in the EDMFT \cite{Smith1,Smith2,Chitra}.
The local moment is coupled to two dissipative baths, one fermionic
and the other bosonic. The bosonic bath describes a fluctuating
magnetic field.
The couplings to the two baths are $J_K$ and $g$, respectively.
The energy dispersions of the two baths are determined self-consistently.
}
\label{fig1}
\end{figure}

\begin{figure}
\caption{
Schematics of a conventional quantum phase transition
in Kondo lattices.
The tuning parameter $\delta$ is the ratio of the RKKY interaction
to the Kondo energy scale $T_K$.
When $\delta$ is small, the local moments are completely screened
by the spins of the conduction electrons, and the system is a paramagnetic
metal. Increasing $\delta$ strengthens the coupling between each local moment
and the fluctuating magnetic field produced by all other local moments, as
measured by the ratio $g/T_K$, where $g$ parametrizes the effective local
problem defined in Fig.~\ref{fig1}b and equation~(\ref{H-imp}). At
$\delta_{\rm loc}^c$,
the corresponding $g/T_K$ would reach the critical value for a phase transition
in the local dissipative problem (see main text); at this point, a local energy
scale $E_{\rm loc}^*$ would vanish.
Increasing $\delta$ also increases the spin susceptibility at the peak
wavevector, $\chi({\bf Q})$, which diverges at a quantum critical point at
$\delta=\delta_c$.
Beyond $\delta_c$, there exists a finite transition temperature
$T_N$ below which the system is an antiferromagnetic metal.
In this conventional case, $\delta_c<\delta_{\rm loc}^c$, i.e.,
$E_{\rm loc}^*$ is finite at the critical point.
The quantum-critical (QC) regime, denoted by the wavy line, is
described by the traditional theory.}
\label{fig2}
\end{figure}

\begin{figure}
\caption{
Schematics of a locally critical
quantum phase transition in Kondo lattices.
The notation is
the same as in Fig.~\ref{fig2}. The crucial new
feature here is that $\delta_c = \delta_{\rm loc}^c$,
i.e., $E_{\rm loc}^*$ vanishes at the quantum critical point.
Local and spatially-extended critical degrees of freedom co-exist
in the quantum critical (QC) regime.}
\label{fig3}
\end{figure}

\newpage

\begin{figure}
\vspace*{2in}
\centering
\vbox{\epsfysize=10cm\epsfbox{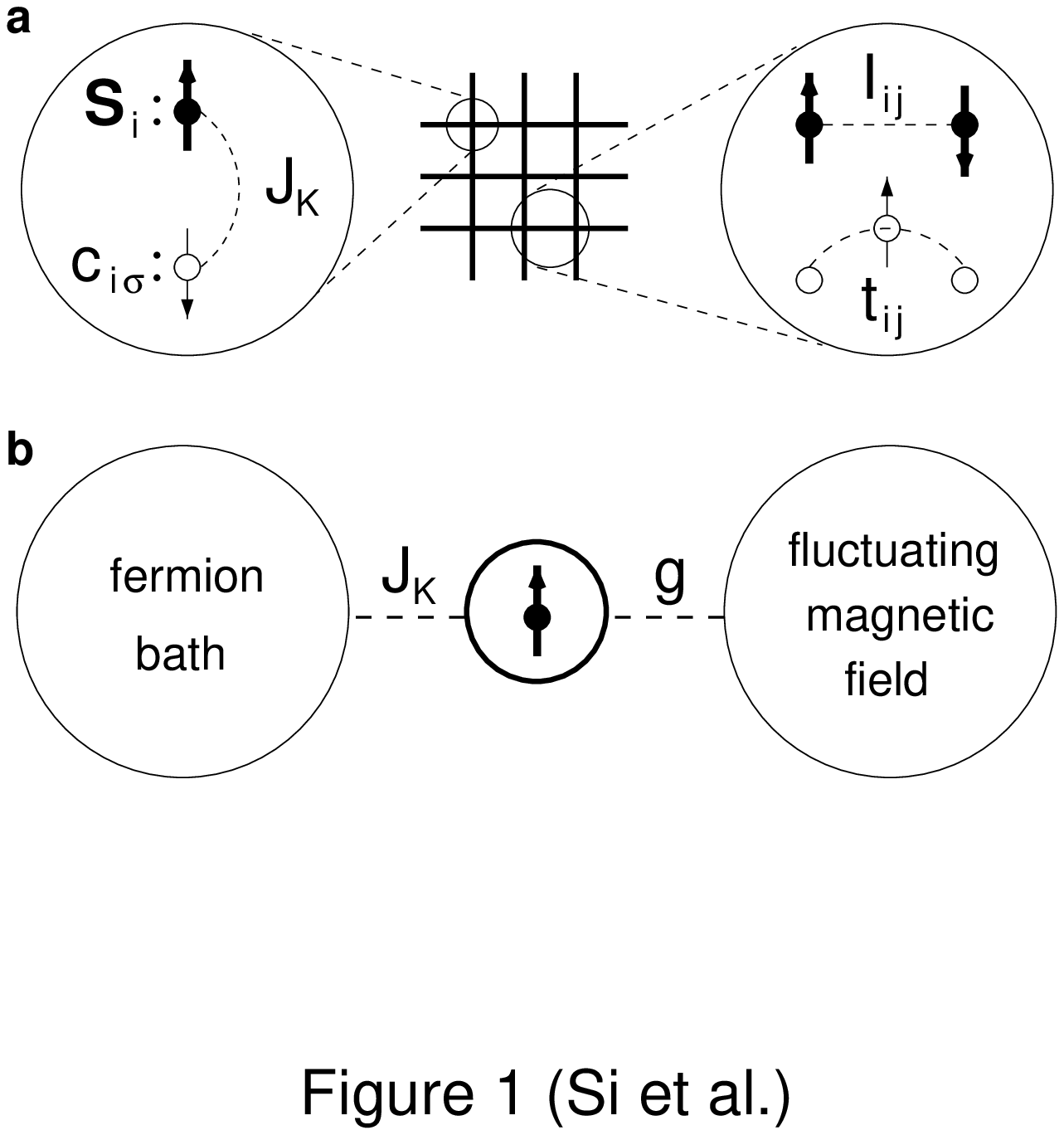}}
\end{figure}

\newpage

\begin{figure}
\vspace*{2.3in}
\centering
\vbox{\epsfxsize=10cm \epsfbox{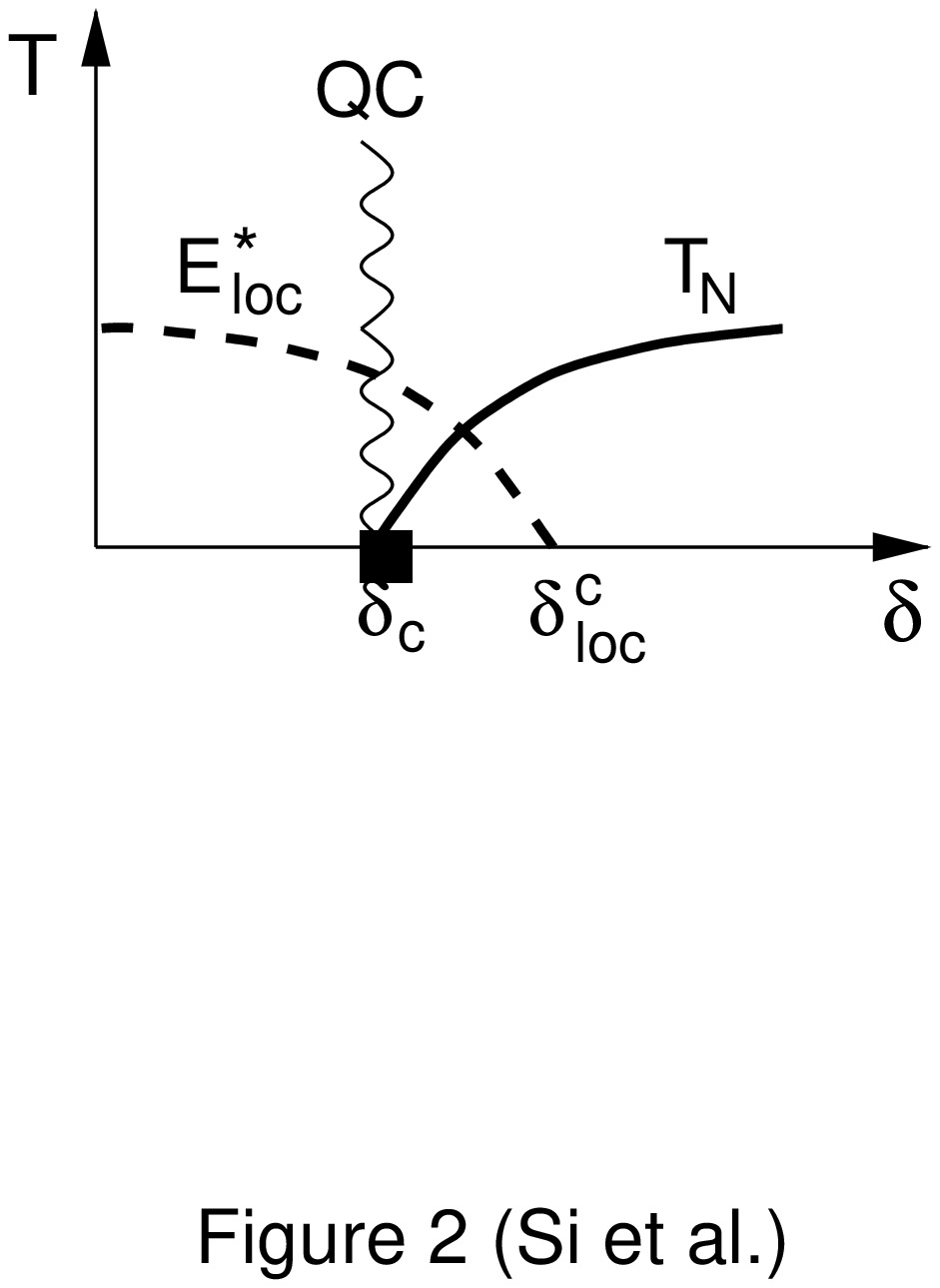}}
\end{figure}

\newpage

\begin{figure}
\vspace*{2.5in}
\centering
\vbox{\epsfxsize=10cm \epsfbox{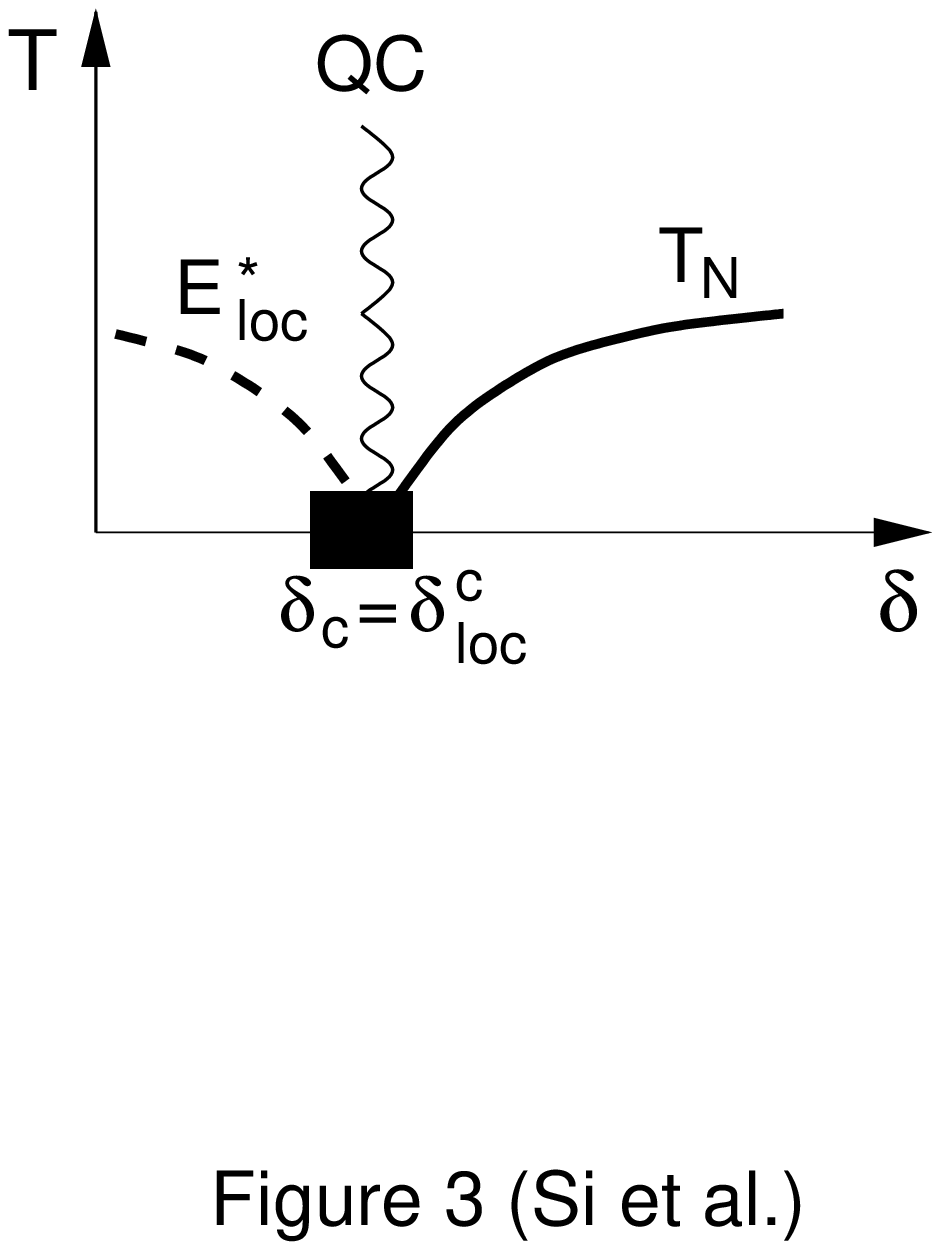}}
\end{figure}


\begin{thebibliography}{9}

\bibitem{Sachdev-book} Sachdev, S.
{\it Quantum Phase Transitions}
(Cambridge Univ.\ Press, Cambridge, 1999).

\bibitem{Maple} Maple, M.\ B.\ {\it et al.}
Non-Fermi-liquid behavior in strongly correlated $f$-electron
materials, J. Low Temp. Phys. {\bf 95}, 225 (1994).

\bibitem{Lonzarich} Mathur, N.\ D.\ {\it et al.}
Magnetically mediated superconductivity in heavy
fermion compounds. {\it Nature} {\bf 394}, 39--43 (1998).

\bibitem{vonLohneysen}
L\"ohneysen, H.\ v.\ {\it et al.}
Non-Fermi-liquid
behavior
in a heavy-fermion alloy at a magnetic
instability. {\it Phys.\ Rev.\ Lett.} {\bf 72}, 3262--3265 (1994).


\bibitem{Steglich} Trovarelli, O.\ {\it et al.}
${\rm Yb Rh_2Si_2}$: Pronounced non-Fermi-liquid effects above
a low-lying magnetic phase transition.
{\it Phys.\ Rev.\ Lett.} {\bf 85}, 626--629 (2000).

\bibitem{Stewart} Heuser, K.\ {\it et al.}
Inducement of non-Fermi-liquid
behavior
with a magnetic field.
{\it Phys.\ Rev.\ B} {\bf 57}, R4198--4201 (1998).

\bibitem{deVisser} Estrela, P., de Visser, A., Naka, T., de Boer, F. R.\ \&
Pereira, L.\ C.\ J. High-pressure study of the non-Fermi liquid
material ${\rm U_2Pt_2In}$. cond-mat/0009324.

\bibitem{RMP} Stewart, G.  Non-Fermi liquid
behavior in d- and f- electron metals.
{\it Rev.\ Mod.\ Phys.}, in press (2001).

\bibitem{Schroder1} Schr\"oder, A., Aeppli, G., Bucher, E.,
Ramazashvili, R.\ \& Coleman, P.
Scaling of magnetic fluctuations near a quantum phase transition.
{\it Phys.\ Rev.\ Lett.} {\bf 80}, 5623--5626 (1998).

\bibitem{Stockert} Stockert, O., L\"ohneysen, H.\ v., Rosch, A.,
Pyka, N.\ \& Loewenhaupt, M.
Two dimensional fluctuations at the quantum critical point of
CeCu$_{6-x}$Au$_x$.
{\it Phys.\ Rev.\ Lett.} {\bf 80}, 5627--5630 (1998).

\bibitem{Schroder2} Schr\"oder, A.\ {\it et al.}
Onset of antiferromagnetism in heavy-fermion metals.
{\it Nature} {\bf 407}, 351--355 (2000).

\bibitem{Lonzarich-review} Lonzarich, G.\ G.
in {\it Electron} (ed.\ Springford, M.) 109--147
(Cambridge Univ.\ Press, Cambridge, 1997).

\bibitem{Hertz} Hertz, J.\ A.
Quantum critical phenomena.
{\it Phys.\ Rev.\ B} {\bf 14}, 1165--1184 (1976).

\bibitem{Millis} Millis, A.\ J.
Effect of a nonzero temperature on quantum critical points in itinerant
fermion systems.
{\it Phys.\ Rev.\ B} {\bf 48}, 7183--7196 (1993).

\bibitem{Coleman} Coleman, P.
Theories of non-Fermi liquid
behavior in heavy fermions.
{\it Physica B} {\bf 259-261}, 353--358 (1999).

\bibitem{SiSmithIngersent} Si, Q., Smith, J.\ L.\ \& Ingersent, K.
Quantum critical behavior in Kondo systems.
{\it Int.\ J.\ Mod.\ Phys.\ B} {\bf 13}, 2331--2342 (1999).

\bibitem{Rosch} Rosch, A.
Interplay of disorder and spin fluctuations in the resistivity near
a quantum critical point.
{\it Phys.\ Rev.\ Lett.} {\bf 82}, 4280--4283 (1999).

\bibitem{Coleman2} Coleman, P.,
P\'{e}pin, C.\ \&
Tsvelik, A.\ M.
Supersymmetric spin operators.
{\it Phys.\ Rev.\ B} {\bf 62}, 3852--3868 (2000).

\bibitem{Leggett} Leggett, A.\ J.\ {\it et al.}
Dynamics of the dissipative two-state system.
{\it Rev.\ Mod.\ Phys.} {\bf 59}, 1--86 (1987).

\bibitem{Hewson} Hewson, A.\ C.
{\it The Kondo Problem to Heavy Fermions}
(Cambridge Univ.\ Press, Cambridge, 1993).

\bibitem{Anderson61} Anderson, P.\ W.\
Localized magnetic states in metals.
{\it Phys. Rev.} {\bf 124}, 41--53 (1961).

\bibitem{Doniach} Doniach, S.
The Kondo lattice and weak antiferromagnetism.
{\it Physica B} {\bf 91}, 231--234 (1977).

\bibitem{Varma} Varma, C.\ M.
Mixed-valence compounds.
{\it Rev.\ Mod.\ Phys.} {\bf 48}, 219--238 (1976).

\bibitem{Smith1} Smith, J.\ L.\ \& Si, Q.
Spatial correlations in dynamical mean-field theory.
{\it Phys.\ Rev.\ B} {\bf 61}, 5184--5192 (2000).

\bibitem{Smith2} Si, Q.\ \& Smith, J.\ L.
Kosterlitz-Thouless transition and short range spatial correlations
in an extended Hubbard model.
{\it Phys.\ Rev.\ Lett.} {\bf 77}, 3391--3394 (1996).

\bibitem{Chitra} Chitra, R.\ \& Kotliar, G.
Effect of Coulomb long-range interactions on the Mott transition.
{\it Phys.\ Rev.\ Lett.} {\bf 84}, 3678--3681 (2000).

\bibitem{Georges} Georges, A., Kotliar, G., Krauth, W.\ \&
Rozenberg, M. J.
Dynamical mean-field theory of strongly correlated fermion
systems and the limit of infinite dimensions,
{\it Rev. Mod. Phys.} {\bf 68}, 13--125 (1996).

\bibitem{Smith3} Smith, J.\ L.\ \& Si, Q.
Non-Fermi liquids in the two-band extended Hubbard model.
{\it Europhys.\ Lett.} {\bf 45}, 228--234 (1999).

\bibitem{Sengupta} Sengupta, A.\ M.
Spin in a fluctuating field: the Bose (+Fermi) Kondo models.
{\it Phys.\ Rev.\ B} {\bf 61}, 4041--4043 (2000).

\bibitem{SachdevYe} Sachdev, S.\ \& Ye, J.
Gapless spin-fluid ground state in a random quantum
Heisenberg magnet.
{\it Phys.\ Rev.\ Lett.} {\bf 70}, 3339--3342 (1993).

\bibitem{Continentino} Continentino, M.\ A.
Universal behavior in heavy fermions.
{\it Phys.\ Rev.\ B} {\bf 47}, 11587--11590 (1993).

\bibitem{Rosch-neutron} Rosch, A., Schr\"oder, A., Stockert, O.\ \&
L\"ohneysen, H.\ v.
Mechanism for the non-Fermi-liquid
behavior
in CeCu$_{6-x}$Au$_x$.
{\it Phys.\ Rev.\ Lett.} {\bf 79}, 159--162 (1997).

\bibitem{Aronson} Aronson, M.\ C. {\it et al.}
Non-Fermi-liquid scaling of the magnetic response in
${\rm U Cu_{5-x}Pd_x~(x=1,1.5)}$.
{\it Phys.\ Rev.\ Lett.} {\bf 75}, 725--728 (1995).

\bibitem{Steglich-priv}
Steglich, F. Private communications (2000).

\bibitem{Vojta} Vojta, M., Buragohain, C.\ \& Sachdev, S.
Quantum impurity dynamics in two-dimensional antiferromagnets
and superconductors.
{\it Phys.\ Rev.\ B} {\bf 61}, 15152--15184 (2000).

\bibitem{Chubukov} Abanov, Ar.\ \& Chubukov, A.\ V.
Spin-fermion model near the quantum critical point: one-loop
renormalization group results.
{\it Phys.\ Rev.\ Lett.} {\bf 84}, 5608--5611 (2000).

\bibitem{Chakravarty}
Chakravarty, S., Halperin, B.\ I.\ \& Nelson, D.\ R.
Two-dimensional quantum Heisenberg antiferromagnet at low temperatures.
{\it Phys.\ Rev.\ B} {\bf 39}, 2344--2371 (1989).

\bibitem{Sachdev} Sachdev, S.
Theory of finite-temperature crossovers near quantum critical points
close to, or above, their upper-critical dimension.
{\it Phys.\ Rev.\ B} {\bf 55}, 142--163 (1997).

\bibitem{Raymond} Raymond, S., Regnault, L.\ P., Flouquet, J.,
Wildes, A. \& Lejay, P.
Pressure dependence of the spin dynamics around a quantum
critical point: an inelastic neutron scattering study of 
${\rm Ce_{0.87}La_{0.13}Ru_2Si_2}$.
cond-mat/0102427.


\bibitem{MFL} Varma, C.\ M., Littlewood, P.\ B., Schmitt-Rink, S.,
Abrahams, E.\ \& Ruckenstein, A.\ E.
Phenomenology of the normal state of Cu-O high temperature superconductors.
{\it Phys.\ Rev.\ Lett.} {\bf 63}, 1996--1999 (1989).

\bibitem{Wilson-Kogut}
Wilson, K.\ G.\ \& Kogut, J.
The renormalization group and the $\epsilon$ expansion.
{\it Phys.\ Rep.\ C} {\bf 12}, 75--200 (1974).

\end{thebibliography}
\end{document}